%% file: main.tex
\documentclass[sigconf]{acmart}

\AtBeginDocument{%
  }

\setcopyright{none}
\copyrightyear{2024}
\acmYear{2024}
\acmDOI{XXXXXXX.XXXXXXX}

\acmConference{ACT3 Technical Whitepaper}
\acmYear{2026}
\setcopyright{none}
\renewcommand\footnotetextcopyrightpermission[1]{}
\usepackage[]{hyperref}  

\input{packages}

\input{macros}

\begin{document}

\title{Architecture Carbon Tool v3: Enabling Sustainability-aware Silicon System Design Exploration}

\author{Vincent T. Lee}
\email{vtlee@meta.com}
\affiliation{%
  \institution{Meta}
  \department{Reality Labs Silicon}
  \city{Seattle}
  \state{Washington}
  \country{USA}
}
\author{Bilge Acun}
\email{acun@meta.com}
\affiliation{%
  \institution{Meta}
  \department{FAIR}
  \city{Menlo Park}
  \state{California}
  \country{USA}
}
\author{Zachary Lewis}
\email{zachlewis@meta.com}
\affiliation{%
  \institution{Meta}
  \department{Reality Labs Sustainability}
  \city{Seattle}
  \state{Washington}
  \country{USA}
}
\author{Carole-Jean Wu}
\email{carolejeanwu@meta.com}
\affiliation{%
  \institution{Meta}
  \department{FAIR}
  \city{Cambridge}
  \state{Massachusetts}
  \country{USA}
}

\begin{abstract}
\input{00-abstract}
\end{abstract}

\maketitle

\input{01-introduction}
\input{02-objectives}
\input{03-modeling}
\input{04-analysis}
\input{05-trends}
\input{06-case-studies}
\input{07-looking-forward}
\input{08-conclusion}

\bibliographystyle{plain} 
\bibliography{references}

\end{document}

%% file: packages.tex
\usepackage{subcaption} 
\usepackage{cleveref}   
\usepackage{listings}
\usepackage{xcolor}
\usepackage{flushend}

\lstdefinelanguage{yaml}{
  keywords={true, false, null, yes, no},
  keywordstyle=\color{blue}\bfseries,
  basicstyle=\ttfamily\small,
  sensitive=false,
  comment=[l]{\#},
  commentstyle=\color{gray}\ttfamily,
  stringstyle=\color{red},
  morestring=[b]',
  morestring=[b]",
  moredelim=[l][\color{orange}]{-\ },
  moredelim=**[l][\color{teal}]{:\ },
}

%% file: 00-abstract.tex
As the carbon cost of manufacturing and operating semiconductor devices has come into sharper focus, sustainability has gradually emerged as a new system architecture design metric.
Like power and performance modeling tools, enabling sustainability-aware silicon systems design and optimizations will require a new generation of electronic design automation and architectural modeling tools.
Towards this end, we present an update to the Architecture Carbon Tool v3 (ACT3) which aims to provide an extensible and customizable modeling platform for research and advanced development to pave the path towards sustainability-aware architectural design space exploration.
Compared to previous versions of ACT, ACT3 provides significantly richer modeling capabilities, enhanced collateral and analysis telemetry, and first order design space exploration capabilities.
This technical brief provides an overview of these expanded capabilities and illustrates ACT3's basic utility across several case studies.
Finally, we identify opportunities for research and development where we expect the research community can contribute towards continuing to improve sustainability modeling and design methodology for silicon systems.

%% file: 01-introduction.tex
\section{Introduction}

The sustainability of silicon systems has become an increasingly important design consideration as the manufacturing costs of advanced semiconductor logic processes, memories, and consumer electronic components have gradually become more measurable.
Like power and performance modeling, enabling sustainability as an emerging metric for silicon systems design and exploration will require new sustainability-focused electronic design automation and architectural modeling tools.
Recently, with the development of advanced silicon manufacturing models like imec.netzero~\cite{imecSSTS}, the research community's visibility to and the tractability of formulating novel sustainability-aware silicon design optimizations has gradually improved.
To further enable sustainability-aware design decisions, the architectural cost of the full system needs to be unified into measurable and reproducible models to understand how different system architectures compare.
This requires unifying carbon emissions models across a wide variety of system components as well as managing and consolidating the mechanics of highly complex systems to get a full system view of the sustainability design trends.

Towards this end, we present an update to the Architecture Carbon Tool (ACT)~\cite{act} which aims to provide a unified, flexible, and extensible architectural modeling platform for research and development of sustainability-aware silicon systems.
Architecture carbon tool v3 (ACT3) provides significantly enhanced modeling capabilities, better collateral and analysis telemetry, and new design space exploration options for first order device carbon emissions projections.
Many of these models are designed with reconfigurability in mind so that sustainability modeling trends, requirements, and modeling data can be improved or adapted to new modeling data as it becomes available.
ACT3 now defines a yaml architecture configuration or spreadsheet bill of materials specification which enables modular composition of components and scalably manages the complexity of large systems (\autoref{sec:bill-of-materials}).
The ACT3 tool also integrates additional component models which have been expanded to better support modeling considerations beyond datacenter class system modeling to wearable class systems (\autoref{sec:models}).

To enhance the analysis process, ACT3 provides new collateral generation capabilities to make results and trends easier to analyze and debug.
ACT3 now emits a static html dashboard export which provides different views and detailed telemetry that can be used to analyze and understand the sustainability bottlenecks in the system.
Since the dashboard is statically rendered, the results are self-contained and can be seamlessly shared with collaborators for analysis.
We have also enhanced ACT3 to generate both yaml manifests and spreadsheet collateral exports to support downstream programmatic and spreadsheet-based modeling workflows.
We expect these quality of life improvements to better facilitate analysis of architectural bottlenecks and sustainability-aware research and development workflows.

ACT3 also introduces several design space exploration capabilities to enable first order trend analysis and projections by annotating the manufacturing year to system components (\autoref{sec:trends}).
Using publicly available technology node data, ACT3 now provides first order technology scaling projection to analyze the impact on the embodied and operational carbon emissions.
ACT3 also integrates the Ember dataset~\cite{ember} which provides the carbon intensity data by year per country; using this data, ACT3 enables trend analysis of the impact of carbon intensity trends over time as well as analyzing the impact of adjusting manufacturing supply chain locations.
We expect these modeling knobs to serve as a foundation for layering more complex design space explorations and analyses.

Finally, we note that ACT3 remains intended to provide rapid first order evaluation to guide design space explorations during the research and development process.
ACT3 is \textit{not} intended to fill the role of life cycle analyses which typically are slower to produce as they require more work to attain the precision required for certification from reviewers and auditors.

%% file: 02-objectives.tex
\section{Key Objectives}

ACT3 is intended to enable first order architectural design space exploration by unifying various modeling capabilities into a scalable framework.

\subsection{Enabling a Unified Full System View}

Production systems are highly complex and composed of a wide variety of components and materials beyond individual silicon systems.
For instance, datacenter class systems consume significant amounts of power (i.e., high operational carbon share) while mobile and wearable systems are battery-powered and have smaller but highly specialized components to reduce weight and satisfy form-factor requirements (i.e., higher embodied carbon share).
To get a complete view of the carbon emissions of such a range of systems, models need to be able to scalably manage and consolidate a comprehensive and full system view across all constituent components.
This requires supporting potentially multiple levels of design hierarchy, and 100s to 1000s of discrete components which contribute to the overall carbon emissions total.
Unifying the individual contributions of each of these components is important because many individual small components may add up and impose an Amdahl's limit with respect to carbon emissions.

Unfortunately, the modeling data is currently fragmented across the literature, online repositories, and datasets, making this full system unified view difficult to construct.
The problem is further exacerbated by the interdisciplinary nature of sustainability-aware silicon system design where some of the modeling data and formulations are beyond the scope of traditional architectural analyses (ex., industrial manufacturing, etc.).
Towards this end, ACT3 aims to provide a hierarchical and extensible yaml representation which is able to serve as a sufficiently generic and scalable architectural specification to capture the diverse range of potential modeling requirements.
ACT3 also aims to be sufficiently extensible to absorb the different range of component models and data so that it can serve as a unified modeling integration point.
Together, this should allow ACT3 to unify and consolidate the wide range of component models to enable a unified full system view for more complete design space exploration.

\subsection{Enabling Design Space Exploration}

ACT3 aims to enable fast first-order design space explorations that help architects identify trends and optimization opportunities early in the design cycle adjacent to other traditional metrics like power, performance, and area.
Sustainability still remains a relatively new design consideration for system design so there are still open questions and opportunities to establish best practices and design optimization methods.
To do this, we need to first make sustainability for different designs measurable and enable design space exploration capabilities to understand the mechanics of sustainability-aware design space exploration.
ACT3 aims to both continue to make sustainability measurable for silicon systems as well as introduce design space exploration capabilities so that designers can more systematically navigate the complex joint trade-offs in modern silicon systems design.
In particular, ACT3 introduces new customizability and parametrization capabilities to design specifications as well as various other design space exploration capabilities to enable more powerful design space exploration capabilities.
We expect that enabling joint modeling over these design considerations will enable exploration over different design considerations more efficiently and enable a more concrete view into the sustainability design tradeoffs.

\subsection{Reproducibility}

Enabling reproducible and transparent modeling results is essential to allow the research community to layer and combine innovation on top of existing work.
Without reproducibility, researchers spend significant time attempting to recreate baseline results rather than advancing the state of the art, and subtle differences in experimental setups can lead to (unintentionally) misleading comparisons.
ACT3 aims to provide a unified modeling codebase and complete experimental setup configuration so that experimental results from potentially different setups and systems can be made reproducible and compared.
By standardizing how experiments are defined, executed, and shared, ACT3 should enable researchers to focus on innovation rather than reproducing and reverse engineering prior work.

\input{bill_of_materials}

\subsection{Extensibility}

Sustainability research for silicon systems continues to evolve and move rapidly due to the accelerated pace of innovation and AI.
To keep up with this rapid pace of change, modeling frameworks need to be made easily extensible to support future-facing updates.
Thus, ACT3 is intended to be reasonably extensible and flexible to adapt to new component models, trend analysis configurations, and modeling data.
ACT3 will provide default values for model configurations, but almost all modeling data can be parametrized to use updated or custom experimental values.
This provides a layer of indirection so that users can plug and play their own (potentially proprietary) scenarios, models, or data.

\subsection{First Order Modeling}

ACT3 is intended to enable rapid first order evaluation to guide design space explorations during the research and development process by trading accuracy for modeling speed.
The first order modeling capabilities are typically sufficient for capturing high level trends from design space explorations and we try to ground ACT3 modeling data as closely as possible on publicly available data to prevent the accuracy from drifting too much.
However, the fidelity and visibility of the data are constantly changing so we limit the modeling objectives for ACT3 to first order projections.
As a result, we note ACT3 is \textit{not} intended for highly precise exercises which may require multiple significant figures of accuracy or formal sustainability audits.
For precise and auditable carbon estimation, which can require access to proprietary databases and information, we defer to more precise life cycle analysis methodologies.
Life cycle analysis requires more detail and analysis of the device bill of materials as technical details such as the material composition and manufacturing process can matter significantly.

%% file: bill_of_materials.tex
\begin{figure*}[t]
  \centering

  \begin{minipage}[t]{0.3\textwidth}
    \centering
    \begin{lstlisting}
name: Sample device
devices:
  soc:
    model: logic
    area: 50 mm2
    process: 5nm
  dram:
    model: dram
    size: 8 GB
  battery:
    model: battery
    capacity: 19.5 Wh
    \end{lstlisting}
    \subcaption{Sample ACT3 Configuration.}
  \end{minipage}%
  \hfill
  \begin{minipage}[t]{0.3\textwidth}
    \centering
    \begin{lstlisting}
name: Macro test
macros:
 AREA: 25 mm2
 PROCESS: 5nm
devices:
 dut:
   model: logic
   area: $(AREA)
   fab_ci: taiwan
   process: $(PROCESS)
    \end{lstlisting}
    \subcaption{Macro Definition Example.}
  \end{minipage}%
  \hfill
  \begin{minipage}[t]{0.3\textwidth}
    \centering
    \begin{lstlisting}
name: MTIA 2i Server
imports:
 # 24 x MTIA2 accelerators per server
 mtia_asic0: mtia2.yaml
 mtia_asic1: mtia2.yaml
 cpu0: epyc9684x.yaml
 cpu1: epyc9684x.yaml
    \end{lstlisting}
    \subcaption{Subsystem Import Example.}
  \end{minipage}

  \caption{ACT3 brings several improvements to the bill of materials file specification syntax to enable scalable and hierarchical definition of subsystems to support complex systems.}
  \label{fig:yaml-comparison}
\end{figure*}

%% file: 03-modeling.tex
\section{Modeling Methodology}

This section provides a summary of modeling methodology updates and additional features added to ACT3.

\subsection{Bill of Materials Specification}
\label{sec:bill-of-materials}

ACT3 introduces an updated bill of materials yaml format specification that enables more flexible and hierarchical definition of silicon system components to scalably manage design complexity (shown in \autoref{fig:yaml-comparison}).
The main body of the yaml file format is the device definitions, which each specify a component in the system.
Each device requires a model type field which specifies which carbon emissions model to use as well as various device-type-specific parameters.
ACT3 will make sure to validate the existence of required device parameters and report any dimensional analysis errors in parameter values; for some parameter values ACT3 will assume a default value if none is specified.

\noindent \textbf{Macro Substitution.} The ACT3 yaml syntax includes support for defining macro substitutions to introduce parametrizability to the bill of materials specification.
The macro substitution is implemented as a pre-processor pass and by default will use the values defined in the bill of materials file.
Default macro definitions can be overwritten at the command line to explore different architectural configurations for design space explorations.
This allows users to reuse models and more scalably propagate configuration and design space exploration parameters across potentially very complex designs.

\noindent \textbf{Hierarchical Subsystem Import.} The bill of materials file supports a recursive import capability to enable imports of subsystem designs.
This is handled by an imports section in the bill of materials which specifies which subsystem bill of materials to instantiate per path.
This allows ACT3 to handle arbitrarily complex and hierarchical systems which may have deep subsystem hierarchies or instantiate a large number of similar component devices.

\noindent \textbf{Dimensional Analysis.} ACT3 integrates the pint~\cite{pint} unit'ing library to handle and enforce dimensional analysis.
All relevant quantities in ACT3 are unit'ed to automatically manage composition of quantities and units over successive calculations.
This effectively eliminates potential modeling bugs due to dimensional analysis errors which are tedious to debug in analytical spreadsheet models.

\noindent \textbf{Manufacturing Year.} ACT3 adds the component manufacturing year to bill of materials components which introduces the notion of time to the model.
Annotating the manufacturing date allows ACT3 to analyze and model historical behavior as well as project future trends for a particular design.
This is valuable for conducting first order projections of industrial trends like technology scaling and carbon intensity changes over time which we will show in \autoref{sec:trends}.

\noindent \textbf{Location.} ACT3 introduces finer-grained specification of the component manufacturing and operating location.
The manufacturing and operating location is important because it enables basic supply chain-aware design space parametrization.
For instance, moving the manufacturing or operating location of a device from one location to another may change the carbon emissions cost due to increased or decreased manufacturing carbon intensity.

\subsection{Component Models}

\label{sec:models}

Each component type in ACT3 is formulated as one of several different types of models: area-based, weight-based, storage capacity-based, energy capacity-based, and operational-based.
The different models supported by ACT3 are shown in \autoref{tab:model_types}; each model will also attribute the carbon emissions by source type such as fabrication, materials, packaging, operation, etc.

\input{model_types}

\subsubsection{Logic Model}

The logic model estimates embodied carbon emissions for semiconductor logic devices based on process node, die area, fabrication yield, and/or gas abatement level.
ACT3 provides several logic models depending on the desired configuration: (1) the legacy original ACT logic model~\cite{act}, (2) a more direct imec.netzero configurable logic model~\cite{imecSSTS}, and (3) an AP logic model also using imec.netzero data but larger AP die sizes~\cite{imecSSTS}.
The legacy logic model computes the carbon per unit area as the sum of three components: energy-related emissions derived from manufacturing energy consumption multiplied by the fab location's carbon intensity, process gas emissions adjusted for the facility's abatement level (supporting 95\%, 97\%, or 99\% abatement), and emissions from raw material consumption (see~\cite{act} for original methodology).
The configurable imec.netzero and application processor (AP) models are based on publicly available nodes on imec.netzero~\cite{imecSSTS} for logic and larger AP die; these configurations can be adjusted as needed by passing a different data configuration to the models.
The total carbon cost in all logic models is then divided by fabrication yield to account for defective dies and supplemented with IC packaging overhead.

\subsubsection{Storage Model}

The storage model largely preserves the original ACT~\cite{act} storage models for estimating the embodied carbon of memory and storage devices such as DRAM, SSDs, and HDDs.
The storage capacity-based model calculates emissions using a carbon-per-gigabyte coefficient specific to each storage process technology and multiplies through by the storage of the specified device.
ACT3 reuses the original ACT modeling data for devices from manufacturers such as SK Hynix, Samsung, Seagate, Western Digital, etc. and scales the estimated carbon by the fabrication yield to again account for any yield losses.

\subsubsection{Capacitor Model}
ACT3 adds a capacitor embodied carbon model based on capacitor type, weight, and carbon intensity depending on the manufacturing location.
Our preliminary model uses data from~\cite{SMITH2018496} for multi-layer ceramic capacitors (MLCC) and tantalum electrolytic capacitors (TEC) which provides projected energy coefficients expressed as energy per unit weight that are multiplied by the fab location's carbon intensity.
The weight-based modeling coefficients can be configured and updated to support other types or modeling values.

\subsubsection{Printed Circuit Board Model}

Most modern consumer and commercial electronic devices require printed circuit board substrates to house components.
These PCBs typically come in several different sizes, numbers of layers, and manufacturing processes which impact the carbon emissions.
ACT3 integrates a preliminary area-based model based on the data from~\cite{Morietal2017} and fits a rudimentary linear model against the number of layers to estimate the carbon cost of a PCB.

\subsubsection{Materials Model}

ACT3 introduces a materials carbon emissions model based on material type and weight used in the system.
This is used to enable modeling of device materials associated with device mechanical support structures such as chassis, frames, or other mechanical components like fasteners.
While these materials may not compose a large fraction of the system carbon emissions, they are still important to track as they can add up with other small components.
Given the wide array of potential components in consumer electronics, we make the materials model configurable to support arbitrary sets of materials and weight coefficients and provide a minimal default configuration.

\subsubsection{Operational Model}

The operational model computes carbon emissions associated with device power consumption using the duty cycle, estimated operational lifetime, and the carbon intensity of the electricity grid at the point of use.
The model uses the Ember~\cite{ember} carbon intensity database to retrieve location and year-specific grid emission factors per location.
This allows ACT3 to account for regional variations in electricity grid carbon efficiency to a first order.

\subsubsection{Battery Model}

Mobile phones and wearable systems, unlike datacenter class systems, operate off a battery.
ACT3 adds a battery model based on energy storage capacity and cathode chemistry (so it multiplies battery capacity by a storage carbon coefficient).
The default basic model in ACT3 provides carbon coefficients for nickel-manganese-cobalt (NMC) cathodes at 87~kgCO$_2$e/kWh and lithium iron phosphate (LFP) cathodes at 61.5~kgCO$_2$e/kWh based on publicly available estimates~\cite{peiseler2024carbon}.
The battery model data is also configurable so that users can add or use their own modeling configurations.

\subsubsection{Manual Model}

ACT3 adds a manual model which enables hardcoded specification of a component's carbon emissions cost; this manual model simply passes the carbon cost specified through to the final result.
This model type is useful for integrating components where the carbon emissions cost is known but the mechanics of the modeling methodology are a black box.
For example, Nvidia provides a high-level life cycle analysis of HGX H100 machines but only provides the total carbon emissions~\cite{h100-lca}.
To integrate this type of modeling data into an ACT3 bill of materials specification, the carbon cost of the HGX100 would be specified as a black box manual model and supplemented with any additional design space exploration components.

\subsection{Carbon Intensity Model}

ACT3 integrates the Ember dataset~\cite{ember}, which provides a carbon intensity estimate of each country or region by year from about 2000 to 2024.
This allows ACT3 to model and explore how carbon emissions trend over time for a particular design or the historical trends based on carbon emissions improvements.
We do note that there are entries in the Ember dataset where the carbon intensity for a particular year and country may be missing or not available.
To address this, ACT3 is programmed to fall back to a linear regression-based projection for any year and country where the data may not be available.
This includes projections for the future beyond years that are currently supported by the Ember dataset, which allows ACT3 to provide a first order projection of future carbon emissions trends.
For examples of how this projection works, we refer readers to \autoref{sec:ci-trends}.

\subsection{Cost (\$) Models}

ACT3 adds both operational cost and carbon emissions offset models to provide a first order estimate of the monetary cost of manufacturing or operating the device.
For operational electricity cost estimates, ACT3 will default to the annual 12-month rolling average for the 2025 U.S. national average industrial cost per kWh~\cite{us_cost_per_kwh} (8.56 cents per kWh).
For carbon emissions credits, ACT3 will default to \$4.80 per metric ton~\cite{carbon_credit}.
However, since the cost for carbon credits varies widely depending on the method, location, etc., we allow the user to parametrize and override these values as needed.

%% file: model_types.tex
\begin{table*}[t]
\centering
\caption{Summary of ACT3 component carbon models, their primary input drivers, and emitted carbon source types.}
\label{tab:model_types}
\small
\begin{tabular}{llll}
\toprule
\textbf{Model} & \textbf{Input Driver} & \textbf{Key Parameters} & \textbf{Emitted Source Types} \\
\midrule
Logic          & Area-based       & Die area, process node, yield, GPA & Fabrication, Packaging \\
IMEC Logic     & Area-based       & Die area, process node, yield       & Fabrication, Packaging \\
AP             & Area-based       & Die area, process node, yield       & Fabrication, Packaging \\
DRAM           & Capacity-based   & Storage capacity (GB)               & Fabrication, Packaging \\
Flash (SSD)    & Capacity-based   & Storage capacity (GB)               & Fabrication, Packaging \\
HDD            & Capacity-based   & Storage capacity (GB)               & Fabrication, Packaging \\
PCB            & Area-based       & Board area, layer count             & Fabrication \\
Battery        & Capacity-based     & Energy capacity (Wh)                & Fabrication \\
Capacitor      & Weight-based     & Component weight, type              & Passives \\
Materials      & Weight-based     & Component weight, material type     & Materials \\
Manual         & Direct           & User-specified carbon (g CO2e)      & User-specified \\
Operational    & Energy-based      & Power, duty cycle, lifetime         & Operation \\
\bottomrule
\end{tabular}
\end{table*}

%% file: 04-analysis.tex
\section{Analysis and Visualization}

\label{sec:analysis}

ACT3 makes several key improvements to the visualization and analysis capabilities to facilitate and accelerate sustainability-aware silicon system design.

\input{carbon_breakdown.tex}

\subsection{Carbon Emissions Breakdown Views}

ACT3 integrates Plotly graph objects~\cite{plotly} to provide different views of carbon emissions breakdowns and system bottlenecks (\autoref{fig:carbon-breakdown}).

\noindent \textbf{Emission Source Type Breakdown.} The source type view (\autoref{fig:carbon-breakdown-source}) partitions carbon emissions by attribution source and distinguishes between fabrication, materials, operational, etc.
This breakdown enables designers to understand where in the manufacturing process carbon costs accumulate in the model.
This view also provides coarse-grained telemetry as to how much the designer may be able to impact the carbon emissions.
For instance, if the operational bottleneck is higher, then the focus should be on power and energy optimization, while if the bottleneck is from fabrication it would indicate that reducing embodied carbon would be more impactful.
The visualization also captures a view of what components of the total carbon emissions cannot realistically be impacted by silicon design optimization such as packaging costs and non-digital logic components.

\noindent \textbf{Device Category Breakdown.} The device category view (\autoref{fig:carbon-breakdown-category}) aggregates emissions by functional device type, providing a high-level view on carbon distribution across different classes of devices (ex., logic and memory).
This abstraction helps architects quickly assess the carbon trade-offs and trends between different device categories without requiring detailed component-level analysis.
For instance, in \autoref{fig:carbon-breakdown-category}, this view shows that logic consumes a large share of the system carbon compared to memory as well as the relative ratio.
This gives us a high-level understanding and full system context view of the potential impact of optimizing memory versus logic on this system which can direct system design optimization efforts.

\noindent \textbf{Subsystem Hierarchy Breakdown.} The subsystem hierarchy view provides a hierarchical decomposition of carbon emissions and bottlenecks across individual device components and device subsystem boundaries.
For example, the subsystem view in \autoref{fig:carbon-breakdown-subsystem} shows the relative carbon emissions by subsystem for an MTIA2 server model (without power) and shares of the different system devices.
In particular, it shows that individually the DRAMs and CPUs consume more embodied carbon per subsystem, but overall the ASICs in the system add up and dominate since there are substantially more of them.

\noindent \textbf{Spreadsheet Collateral.} ACT3 will export a spreadsheet view of the resulting carbon emissions breakdown by component to the output collateral directory.
This allows the results generated by ACT3 to be usable or merged into existing spreadsheet models or workflows for downstream analysis.

\subsection{Delta Analysis}

ACT3 enables a delta analysis visualization view that allows designers to compare and contrast the sustainability of different design variants or models.
This precise comparison view is important for making trends explainable and evaluating how much better one design point may be relative to another.

\begin{figure*}
\centering
\includegraphics[width=\linewidth]{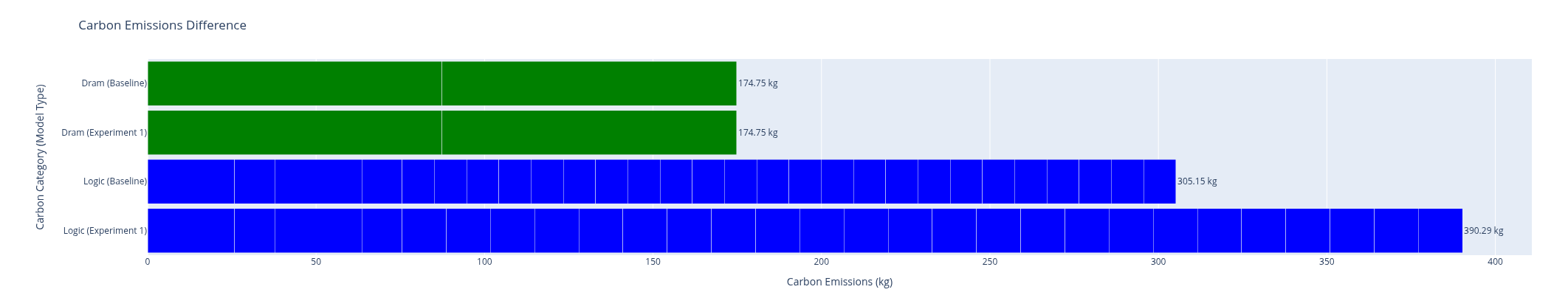}
\caption{Sample delta analysis comparison view between first order MTIA2 and MTIA server model based on~\cite{mtia2, mtia}. The delta analysis view enables a view of how design decisions shift the carbon emissions across the system.}
\label{fig:delta_analysis}
\end{figure*}

To illustrate the utility of the delta analysis comparison, \autoref{fig:delta_analysis} shows a hypothetical comparison of first order MTIA and MTIA2 server models considering compute and memory only.
At a high level, the delta visualization allows for an efficient comparison of where the differences in carbon emissions contributions are in the system to answer questions such as "what is the first order impact on sustainability across MTIA chip generations?"
The delta visualization also provides a coarse-grained attribution of where the carbon emissions are different; in this example, we see that the only difference in the bill of materials is the compute chip parameters (ex., area, etc.) which aligns with the experimental setup.
In addition, ACT3 will also provide a detailed table view of the differences across the two designs which, for significantly more complex systems, can help analyze more precisely where the differences between the designs occur.

\subsection{Optimization Metrics}

ACT3 integrates and computes several carbon-aware optimization metrics from prior work~\cite{carbon_efficient_design}.
These include carbon-delay product, carbon-energy product, carbon energy-squared product, and carbon-squared energy product.
Based on the carbon emissions results and operating time/power usage, ACT3 will calculate each of these metrics for the model which can be passed to downstream design optimization workflows.
These metrics are also exported to the result manifest assets such as the yaml result dump or spreadsheet export which can be used for downstream workflows such as optimization.

%% file: carbon_breakdown.tex
\begin{figure*}[t]
    \centering
    \begin{subfigure}[b]{0.32\textwidth}
        \centering
        \includegraphics[width=\textwidth]{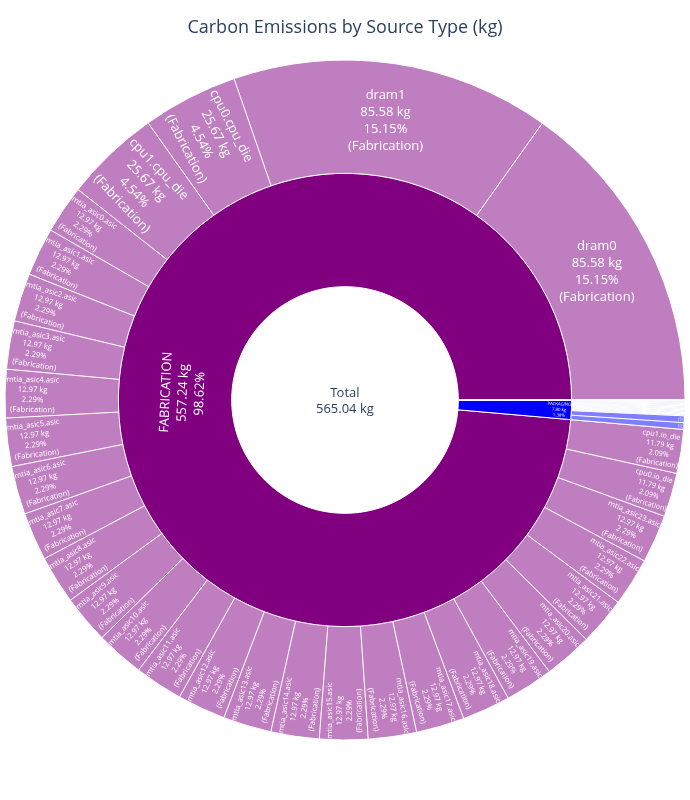}
        \caption{By carbon emissions source type.}
        \label{fig:carbon-breakdown-source}
    \end{subfigure}
    \hfill
    \begin{subfigure}[b]{0.32\textwidth}
        \centering
        \includegraphics[width=\textwidth]{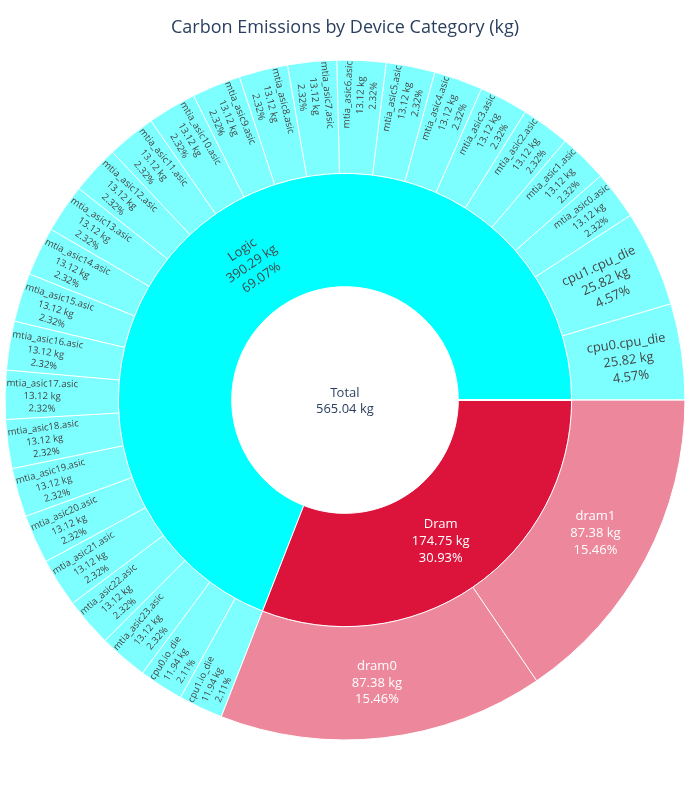}
        \caption{By device category type.}
        \label{fig:carbon-breakdown-category}
    \end{subfigure}
    \hfill
    \begin{subfigure}[b]{0.32\textwidth}
        \centering
        \includegraphics[width=\textwidth]{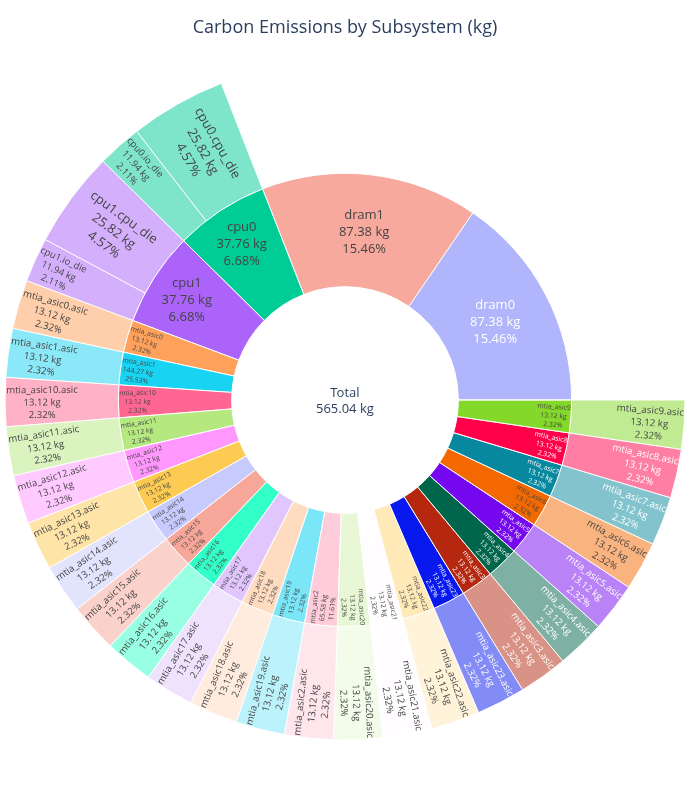}
        \caption{By device subsystem hierarchy.}
        \label{fig:carbon-breakdown-subsystem}
    \end{subfigure}
    \caption{Sample first order carbon emissions estimate for MTIA2 server model based on~\cite{mtia2}. ACT3 provides a full system visualization of the carbon emissions telemetry based on the source type, device category type, or device subsystem hierarchy, providing different views of the system bottlenecks.}
    \label{fig:carbon-breakdown}
\end{figure*}

%% file: 05-trends.tex
\section{Design Trends and Exploration}

\label{sec:trends}

The sample carbon emissions results presented in \autoref{sec:analysis} represent a snapshot in time using assumptions about the state of silicon and manufacturing technologies today.
However, we expect that as silicon systems and devices mature, the manufacturing costs of various materials will shift around and exacerbate or reduce bottlenecks.

\subsection{Technology Scaling Trends}

We expect that silicon technology will continue to improve, resulting in higher carbon intensity per unit area with more advanced and smaller technology processes.
To project the first order impact of technology scaling, ACT3 provides a technology scaling estimation based on the publicly available power scaling data~\cite{tsmc65_40, tsmc45_40, tsmc20_16, tsmc10_7, tsmc7_7euv, tsmc7_5, tsmc3_2, tsmc2_a14}.
The power scaling factors between process generations are successively compounded to project the scaling factors between each of the supported technology processes.
We do note that there are some process technology node generations where public data is missing (ex., TSMC 16nm to TSMC 10nm); to fill in these scale factors, we use a conservative 0.8x process scaling factor between generations.

\begin{figure}[t]
\centering
\includegraphics[width=\linewidth]{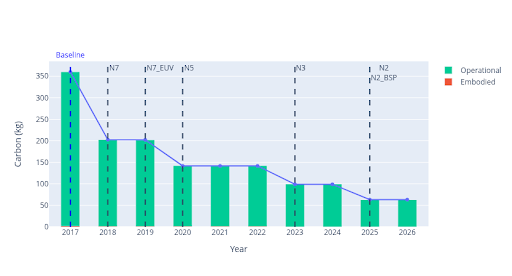}
\caption{Sample ACT3 technology scaling trend analysis. ACT3 uses publicly available first order technology scaling to model power scaling trends which reduce carbon emissions in successive generations of process technologies.}
\label{fig:act3_tech_scaling}
\end{figure}

An example of this technology scaling trend analysis is shown in \autoref{fig:act3_tech_scaling}.
It should be noted that ACT3 will both apply the operational power scaling analysis as well as a first order quadratic area reduction to logic devices (i.e., 7nm to 5nm scaling will reduce logic area by 25/49$\times$); this results in both an operational and embodied carbon emissions reduction.
We note that in practice, the precise technology power and area scaling behavior of silicon will vary depending on the specific silicon device, manufacturer, and power types (ex., dynamic, leakage, logic versus memory, etc.) so the estimates should be treated as first order projections.
Finally, we note that ACT3 can be configured to apply custom technology scaling factors if different and more precise scaling factors are desired.

\subsection{Carbon Intensity Trends}
\label{sec:ci-trends}

We expect the carbon intensity of different countries to shift over time depending on environmental policy, innovations in manufacturing efficiency, and other technological trends.
To model how carbon intensity trends over time impact the sustainability of silicon systems, ACT3 integrates the Ember dataset~\cite{ember}.
The Ember dataset contains the estimated carbon intensity of electricity generation across different regions over the last two decades.
This allows us to model the historical carbon intensity per country as well as project high-level trends into the future.

\begin{figure}[t]
\centering
\includegraphics[width=\linewidth]{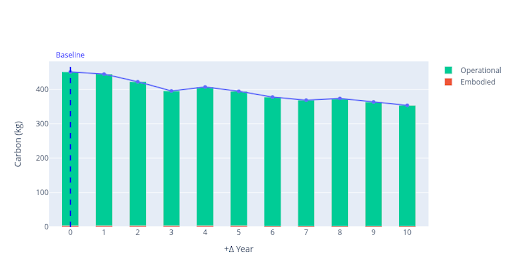}
\caption{Sample ACT3 carbon intensity trend analysis over time. ACT3 uses the Ember dataset to model and project the carbon intensity trends and impact on a design over time.}
\label{fig:ci_dse}
\end{figure}

A sample analysis of how the carbon intensity impacts the sustainability of a design is shown in \autoref{fig:ci_dse}.
In this example, the carbon intensity directly impacts the operational emissions as more sustainable carbon intensity levels reduce the carbon emissions from energy sources.
For countries and/or years where the Ember dataset does not directly provide carbon intensity data, ACT3 will construct a linear regression model and extrapolate the carbon intensity.

\subsection{Supply Chain Trends}

ACT3 introduces the basic capability to experiment with supply chain manufacturing location changes to explore the impact on embodied emissions.
To do this, ACT3 first computes the baseline carbon emissions for each component in the system.
ACT3 then takes a scaling configuration which specifies which subsystem paths or individual devices to relocate, and what manufacturing location to move them to.
ACT3 estimates the relocated manufacturing cost by scaling the baseline cost by the ratio of the carbon intensity of the new location to that of the original location.

\input{supply_chain}

To illustrate this capability, we use the Dell R740 server~\cite{dellr740lca} as a baseline and evaluate the impact of relocating all component manufacturing (CPUs, DRAM, and SSDs) from Taiwan to three hypothetical alternative locations: South Korea, Japan, and the United States.
For each scenario, the same scaling configuration is applied uniformly to all device paths in the bill of materials, ensuring that the entire supply chain is relocated together.
\autoref{tab:supply_chain} shows the estimated total carbon emissions for each manufacturing location.
While this is a simple example, for more complex configurations we expect this capability to more scalably answer "what if" supply chain questions.

%% file: supply_chain.tex
\begin{table}[b]
\centering
\caption{Sample supply chain analysis for Dell R740 server with manufacturing relocated to different countries.}
\label{tab:supply_chain}
\small
\begin{tabular}{lrrr}
\toprule
\textbf{Manufacturing Location} & \textbf{Carbon (kg)} & \textbf{$\Delta$ (kg)} & \textbf{$\Delta$ (\%)} \\
\midrule
Taiwan (Baseline) & 1518.2 & --- & --- \\
South Korea       & 1018.0 & $-$500.1 & $-$32.9\% \\
Japan             & 1162.9 & $-$355.3 & $-$23.4\% \\
United States     &  925.5 & $-$592.6 & $-$39.0\% \\
\bottomrule
\end{tabular}
\end{table}

%% file: 06-case-studies.tex
\section{Case Studies}

This section shows how ACT3 can be used to conduct first order modeling for analyzing trends in modern silicon systems and sustainability-aware system design space exploration.

\subsection{Analyzing Current Trends}
\label{sec:analyzing-trends}

One of the key historical challenges in silicon research is the sheer volume and speed at which the aggregate body of research is expanding.
Reading through and extracting relevant information manually generally does not scale beyond a few dozen or hundred papers for a single researcher at a time and/or requires significant manual effort.
To bridge this gap, we use AI-assisted automation to review conference proceedings so that we can use ACT3 to enable a view into current sustainability trends in silicon research; by leveraging AI, it makes these types of high-level trend analyses over large bodies of work significantly more scalable and real time.

To demonstrate the efficacy and utility of this approach, we take the proceedings from ISSCC 2024-2026 to generate a first cut set of ACT3 specifications from more than several 100 cutting-edge silicon research papers.
To do this, we extract text from each paper in the proceedings and use AI tools to extract the chip specifications and generate an ACT3 bill of materials.
Most papers published at ISSCC have a chip specification and/or table of results which provides the necessary and sufficient modeling data to generate a first order carbon emissions estimate with ACT3 (e.g., technology node, die area, power, energy/throughput, etc.).

To quality control and validate the results, we randomly sample and cross-check 25/105 (\~25\%) of the valid generated bills of materials for ISSCC 2026 against the original papers to evaluate how accurate the extraction procedure is.
Of the papers we reviewed manually, we found that the AI automation made a mistake in one paper within the sampled set (less than 5\% error rate)\footnote{We used Claude 3.6, but we expect accuracy to be better with newer models.}.
We note that the extraction process fails for invited or plenary papers that are not about a specific chip, as well as papers where chip specifications are embedded in chip micrograph images that are not captured by text-based extraction.
Finally, we also conduct several quality control passes using the AI to specifically identify issues such as missing fields or flag and correct mistakes in the inferred assumptions.

\begin{figure}[b]
\centering
\includegraphics[width=\linewidth]{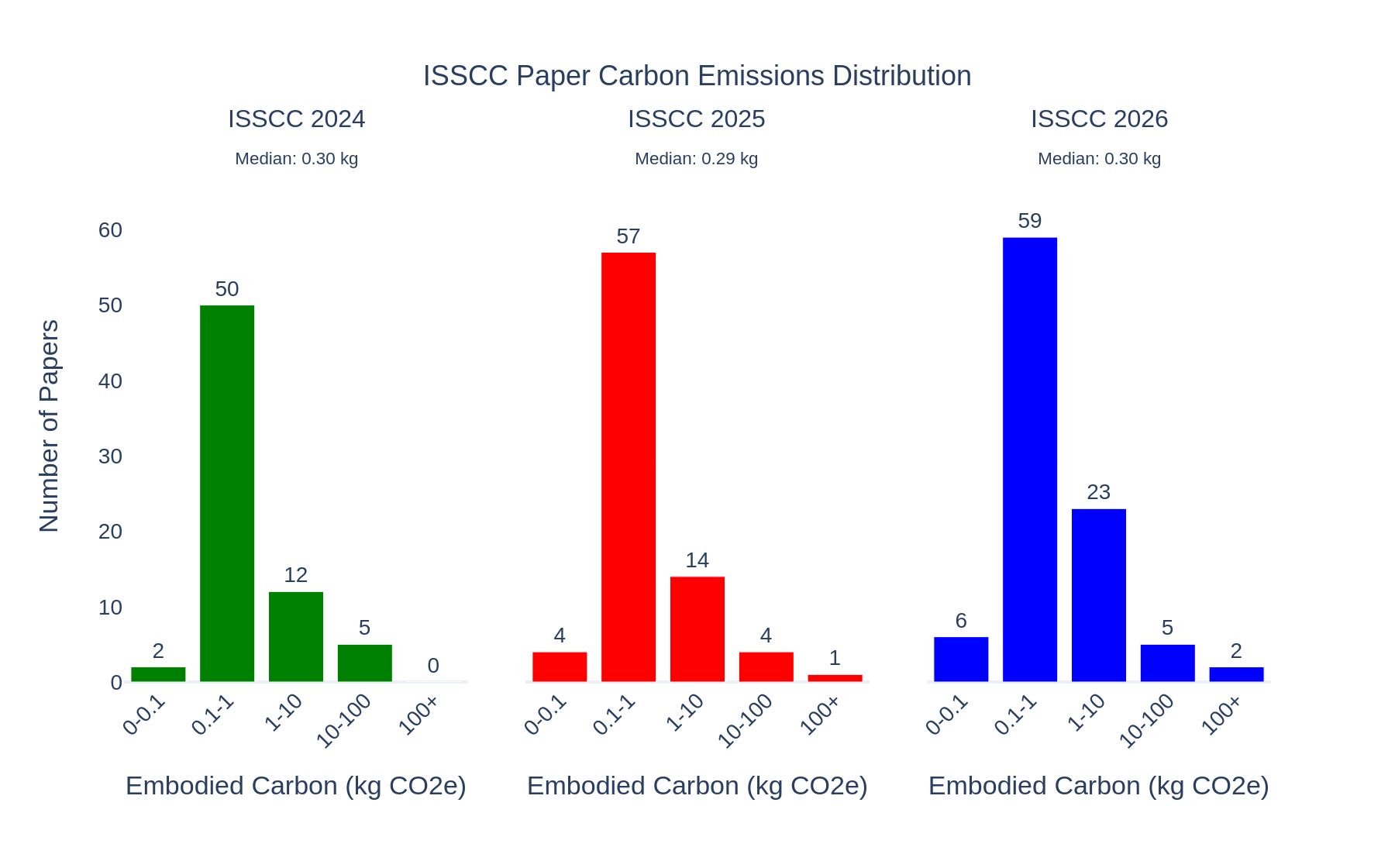}
\caption{Distribution of estimated embodied carbon emissions (kg CO2e) for a subsample of chips presented at ISSCC 2024-2026.}
\label{fig:isscc_cdf}
\end{figure}

\autoref{fig:isscc_cdf} shows a histogram distribution of the estimated embodied carbon emissions for extracted papers across ISSCC 2024, ISSCC 2025, and ISSCC 2026; we focus on embodied emissions since many papers report normalized energy per throughput metrics rather than direct power values.
The distribution of results spans several orders of magnitude, which largely reflects the diversity of chip design sizes presented at ISSCC ranging from small-scale sensor SoCs to large AI accelerators, which roughly correlate with embodied carbon costs.
Over time the data shows a very weak shift over the last three years to slightly higher emissions, but more complete data over the years is necessary to validate the trends since technology process changes shift slowly; we leave more of this analysis for future work.

At a higher level, the results demonstrate how we can begin to harness some of the cutting-edge results from VLSI research more efficiently and at scale for first order carbon modeling trends.
Each conference batch in this workflow took roughly 8 hours to process using a relatively unoptimized workflow which means that with properly prepared infrastructure these types of cutting edge research analyses can be conducted in almost real time (i.e., as research results are made available).
Finally, the extracted specifications from this type of workflow can be used to feed downstream design space explorations and potentially enable new design space exploration and research directions.

\subsection{Server Architecture Exploration}
\label{sec:server-dse}

\begin{figure*}[t]
\centering
\includegraphics[width=\linewidth]{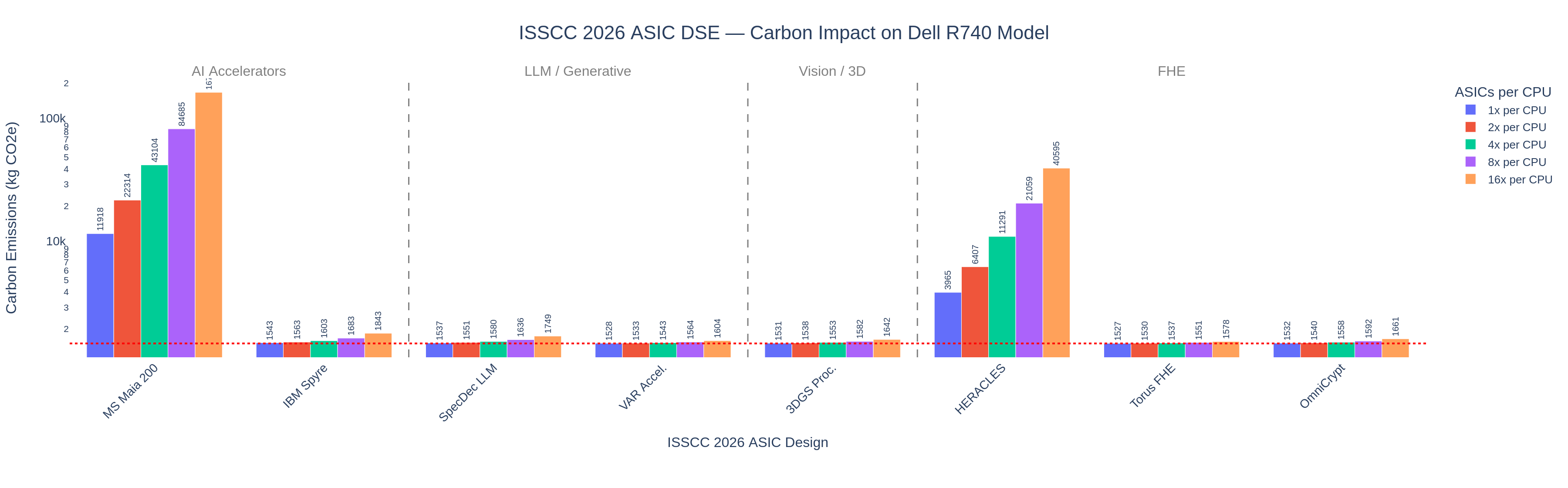}
\caption{Carbon emissions impact on Dell R740 server augmented with selected ISSCC 2026 ASIC coprocessors at 1, 2, 4, 8, and 16 instances per CPU. Smaller accelerators have limited impact on overall carbon emissions while large-die accelerators (Maia 200, HERACLES) quickly dominate system carbon emissions.}
\label{fig:isscc_asics_dse}
\end{figure*}

To illustrate the extensibility and scalable design space exploration capabilities of ACT3, we provide a sample design space exploration using the Dell R740 server as a base architecture.
The Dell R740 architecture is a well-documented (albeit older) 2U rack server whose lifecycle assessment data is publicly available~\cite{dellr740lca} and was studied in the original ACT implementation.
We refactor and parametrize the Dell R740 server specification to illustrate how ACT3 allows for scalable design space exploration of different system design configurations.
In particular, we explore configurations from eight different ASICs spread across AI~\cite{xu2026maia, cohen2026spyre}, LLM and generative acceleration~\cite{wang2026specdec, yue2026var}, Vision~\cite{wang2026threedgs}, and homomorphic encryption~\cite{golder2026heracles, yu2026torusfhe, putra2026omnicrypt} use cases from ISSCC 2026 to explore the potential impact of different cutting-edge ASIC coprocessors on the overall system emissions.
In this experiment, we hold all other server components fixed (CPUs, SSDs, and DRAM modules) and vary the number of ASIC coprocessors per CPU at 1, 2, 4, 8, and 16.
The results are shown in \autoref{fig:isscc_asics_dse}.

At a high level, we observe that in many cases the proposed works do not substantially impact the overall embodied carbon emissions aside from large ASICs like MAIA~\cite{xu2026maia} (which also adds additional memory to the model) and HERACLES, which also have substantially more advanced process technology nodes (3nm) and fairly large die sizes.
More importantly, the modeling results illustrate the value of evaluating these sorts of architecture design decisions in full system context to understand at what point the architectural modifications begin to meaningfully impact carbon emissions.
This is because as different components are optimized, added, or removed from such a system, the Amdahl's limitation for carbon emissions will shift around which in turn impacts where to focus design optimization efforts.
As a result, system designers will need to continuously monitor the full system view to evaluate whether various sustainability-focused optimizations are materially impactful on overall system sustainability.

%% file: 07-looking-forward.tex
\section{Looking Forward}
\label{sec:looking-forward}

We expect that ACT3 will continue to facilitate carbon-aware silicon systems design, but there are still some gaps and opportunities towards enabling a complete unified sustainability modeling tool.

\subsection{AI-assisted Sustainability Analyses}

Like other interdisciplinary architecture research domains, bridging the gap between silicon system design and sustainability has historically been challenging due to the dual expertise nature of the problem space.
To bridge this gap, we expect advancements in AI to enable us to capture, encode, and more scalably deploy sustainability expertise via AI-assisted automation by raising the abstraction level.
Advanced AI in particular is exceedingly powerful at knowledge summarization and code generation, but still needs tools like ACT3 for complex deterministic modeling calculations.
By pairing AI-assisted automation with tools like ACT3, it should be possible to more efficiently digest large volumes of sustainability work into usable telemetry and models for sustainability-aware silicon systems design.
This empowers silicon systems architects to focus on system optimizations and abstract away tedious model and design transcription workflows as well as the underlying complexities and expertise requirements of sustainability research.
We expect these types of AI-assisted automations to more easily democratize the research process in sustainability-aware silicon systems design.

\subsection{Gaps in Carbon Modeling Data}

ACT3 attempts to consolidate data that is publicly available across various fragmented sources as much as possible, but there is still a long way to go.
In particular, there remains a wide range of consumer electronics and components where carbon modeling data is either proprietary, preliminary, or non-existent.
In order to enable more accurate and complete system architecture models, we will have to continue to expand the available set of system components supported by carbon modeling tools to support the diverse landscape of consumer and commercial devices.
For instance, embedded and wearable systems have comparatively smaller component sizes and diverse components (ex., flexible PCBs, lighter materials, lenses, etc.) compared to better-studied datacenter class systems.
We expect that going forward, diversifying carbon modeling capabilities to encompass the wide range of devices will become increasingly important, especially as we start to see emerging and highly complex wearable AI systems.
Towards this end, ACT3 attempts to provide sufficient extensibility so that when better modeling data or changes are made available, it can more seamlessly absorb these updates.

\subsection{Towards a Device Model Library}

The consumer and commercial device and component landscape is enormous and cannot be fully covered by a single individual team.
As a result, to enable a consolidated view across the industry, we will need to continue to aggregate device models into a library to support research and development across the community.
We expect that over time, gradual consolidation and accumulation of publicly available sustainability models for silicon systems will enable new and more meaningful architectural research capabilities and evaluation targets for new design methodologies.
Increasing the size and scale of such a model library will help promote formulation of sustainability-aware benchmarking data and provide a testbed to understand how well sustainability-aware optimizations generalize across silicon systems.
Over time, these will help enable new insights into design trends and how we formulate sustainability-aware silicon systems design and optimization moving forward.

\subsection{Sustainability-aware Design Optimization}

One of the key opportunities for sustainability-aware silicon systems design is new sustainability design optimization approaches and rules of thumb that are complementary to existing power, performance, and area optimizations and requirements.
Recent work has proposed a collection of optimizations for datacenter class systems like in~\cite{carbon_explorer} and~\cite{google_tpu_lca}; however, sustainability-aware optimizations for other consumer systems like mobile devices and wearable systems remain less well-established and explored.
To meaningfully impact the design methodology, sustainability will (1) need to be integrated into the silicon design metrics and (2) the tools will need to recommend optimizations that are complementary to the design constraints of production systems.
Wearable and mobile systems in particular are subject to stricter power, energy, and battery life constraints which impose a different range of architectural design considerations and decisions.
For example, many of these design considerations are (loosely) correlated (ex., area and performance, weight, form-factor, and energy budget); as a result, there is significant opportunity to understand how these considerations jointly interact and behave so that we can then formulate new joint sustainability optimization techniques.

%% file: 08-conclusion.tex
\section{Conclusion}

We presented ACT3, which updates the Architecture Carbon Tool~\cite{act} and provides a unified, scalable, and configurable first order modeling platform for sustainability-aware silicon system design exploration.
ACT3 integrates an enhanced scalable bill of materials specification, expanded and flexible component models, and integrated design space exploration and trend analysis capabilities for sustainability-aware silicon systems design.
ACT3 lowers the barrier for designers to explore the environmental impact of design decisions alongside traditional considerations such as power, performance, and area.
These added capabilities allow ACT3 to provide a foundation upon which the industry and research communities can formulate increasingly sophisticated sustainability-aware design methodologies, carbon models, and system models.
Looking forward, we expect ACT3 and similar tools to continue to drive the transition toward a more sustainable semiconductor industry.